\documentclass[11pt]{article}
\usepackage{graphicx} 
\usepackage[a4paper]{geometry}
\RequirePackage{amsthm,amsmath,amsfonts,amssymb}
\usepackage[sort]{natbib}
\usepackage{url} 
\usepackage{algpseudocode}
\usepackage[linesnumbered,ruled,vlined]{algorithm2e}
\usepackage[
  colorlinks=true,
  linkcolor=blue,
  citecolor=blue,
  urlcolor=blue
]{hyperref}
\usepackage{placeins}
\usepackage{float}
\usepackage{fullpage}
\usepackage{tikz}
\usetikzlibrary{positioning, quotes, trees, shapes.geometric}
\usepackage{standalone}
\usepackage{comment}
\usepackage{subcaption}
\usepackage{booktabs}
\usepackage{xcolor}
\usepackage{multirow}
\usepackage{booktabs}
\usepackage{makecell}
\usepackage{csquotes}
\usepackage{titlesec}

\begin{document}

\title{\textbf{Data generation for the logistic regression model under controlled conditions}}


\author{
Kabiru Abubakari\textsuperscript{1,3} \& 
Silvia Liverani\textsuperscript{1}
}

\date{}

\newcommand{\affiliations}{%
\vspace{-0.5cm}
\noindent
\textsuperscript{1}School of Mathematical Sciences, Queen Mary University of London, United Kingdom\\
\textsuperscript{3}Royal Botanic Gardens, Kew, Richmond, Surrey, United Kingdom\\
}

\maketitle
\affiliations
\begin{abstract}

Generating data with desired properties for a logistic regression model under controlled conditions is challenging. We propose an integrated iterative bisection algorithm that determines values of the intercept and regression coefficients capable of generating data with a specified prevalence and predictive accuracy. The performance of the proposed method is demonstrated through simulation studies. We further show that the choice of parameter values, or, in a Bayesian setting, the choice of priors for these parameters, induces a particular level of predictive accuracy. This relationship should therefore be taken into account when generating data for model evaluation under controlled conditions.

\end{abstract}

\hspace{0.45cm}\textbf{Keywords:} Bisection algorithm, data generation, logistic regression, prevalence

\section{Introduction}\label{sec:intro}
The logistic regression model is widely used to analyse the relationship between a set of explanatory variables and a binary response, and is also a fundamental supervised learning method in machine learning. Owing to its interpretability and ease of implementation, it has become the standard analytical tool across a wide range of disciplines, including epidemiology, engineering, economics, and ecology. In addition to estimating the effects of explanatory variables on a binary outcome, logistic regression can also be used to predict the probability that future observations belong to a particular class.

Simulation studies often require datasets with prespecified prevalence and predictive performance in order to evaluate and compare statistical methods under controlled conditions. However, generating data that simultaneously satisfy both requirements is challenging because the prevalence and discriminatory ability of a logistic regression model are jointly determined by its parameters.

Predictive performance is commonly assessed using measures derived from the receiver operating characteristic (ROC) curve, most notably the area under the curve (AUC) \citep{lieli2019using}. Logistic regression produces fitted values that can be interpreted as the predicted probabilities of the outcome of interest, such as the risk of a disease or the presence of a species at a particular location. To classify observations into positive or negative outcomes, a threshold (or cutoff) is applied to these predicted probabilities. For a given cutoff value, observations with predicted probabilities exceeding the threshold are classified as positive, whereas those with predicted probabilities below the threshold are classified as negative \citep{austin2012interpreting}. By varying the cutoff across the range of possible values, the ROC curve summarizes the trade-off between sensitivity and specificity and provides a comprehensive assessment of the model's discriminatory ability. 

The choice of cutoff value induces a trade-off between true positive and false positive classifications. Consequently, the receiver operating characteristic (ROC) curve summarizes the relationship between the true positive rate (TPR) and the false positive rate (FPR) across a range of cutoff values \citep{Lydia2006using,lieli2019using}. The TPR, also known as sensitivity, is the proportion of positive outcomes that are correctly classified by the model, whereas the FPR is the proportion of negative outcomes that are incorrectly classified as positive. Because binary classification is subject to two types of errors, namely false positives (FPs) and false negatives (FNs), a perfect classifier would achieve a TPR of 1 and an FPR of 0. Accordingly, good predictive models are characterized by TPR and FPR values that are close to these ideal limits, reflecting a high ability to distinguish between positive and negative outcomes \citep{pepe2004limitations}.

The relationship between predictor coefficients and model discrimination, typically measured by the area under the receiver operating characteristic curve (AUC), has been widely studied. \citet{Danping2013} highlighted that predictive accuracy in ROC analysis depends on the effects of the predictors, indicating a direct relationship between regression coefficients and model discrimination. For the case of normally distributed predictors, \citet{austin2012interpreting} examined how the coefficients of a logistic regression model determine the resulting AUC. Likewise, \citet{dodd2003semiparametric} developed an AUC regression framework that quantifies the effect of predictor changes on model discrimination. Together, these studies show that the magnitude of the regression coefficients plays a key role in determining the predictive performance of logistic regression and related classification methods. 

This relationship has important implications for simulation studies in which data are generated to exhibit predefined characteristics. When binary outcomes are simulated from a logistic regression model, the choice of coefficient values affects not only the prevalence of the outcome but also the achievable level of predictive accuracy. Consequently, the relationship between coefficients and discrimination should be taken into account when generating data with desired predictive properties. In a Bayesian setting, an analogous issue arises through the specification of prior distributions for the regression coefficients. Priors on effect sizes induce corresponding prior distributions on predicted probabilities \citep{newman2025logistic}. For example, priors that favour large effect sizes can produce predicted probabilities concentrated near 0 or 1, leading to highly separable classes and AUC values close to 1. Thus, the choice of priors influences both parameter estimation and the predictive characteristics of the generated data. Despite the importance of controlling both prevalence and predictive accuracy, existing approaches do not provide a straightforward way to target both characteristics simultaneously.

In this paper, we propose an iterative bisection algorithm for identifying logistic regression parameter values that generate data with both a target prevalence and a target level of predictive accuracy simultaneously. Our work extends that of \citet{austin2023iterative}, who developed separate iterative bisection algorithms for generating data with either a target prevalence or a target level of accuracy, but not both simultaneously. Applying these algorithms sequentially does not generally achieve both objectives. To address this limitation, we integrate the two approaches into a single algorithm that jointly searches the parameter space for values satisfying both targets. This allows the properties of the simulated data, including prevalence and predictive accuracy, to be controlled explicitly.

The remainder of the paper is organised as follows. Section~\ref{sec:logistic} introduces the logistic regression model, while Section~\ref{sec:roc} describes measures of classification performance. Section~\ref{sec:algorithm} presents the proposed algorithms for generating data from a logistic regression model under controlled conditions. Simulation results are reported in Section~\ref{sec:sim}, and the distributions of the predicted probabilities are examined in Section~\ref{sec:dist_predicted_prob}. Finally, Section~\ref{sec:disc} concludes with a discussion of the findings and their implications.
\section{The logistic regression model}\label{sec:logistic}

The logistic regression model uses the logit link function, that is, the sigmoid function $\sigma: \mathbb{R} \to \left(0, 1 \right)$.
This model is used when the response is binary, that is $Y_i \in \left\{0, 1\right\}$ with available explanatory variables \(\mathbf{x}\) consisting of predictors $\text{x}_1, \text{x}_2, \dotsc, \text{x}_k$, and $\boldsymbol{\beta} = \left(\beta_0, \dotsc, \beta_k \right)$. The logistic regression model can be stated in terms of the probability of $Y_i=1$ conditioned on \(\mathbf{x}\) as follows
\begin{equation}
    \eta(\mathbf{x}_i) = \beta_0 + \beta_1\text{x}_1 + \beta_2\text{x}_2 + \dotsc + \beta_p\text{x}_p
    \label{eq:eq2.1}
\end{equation}
\begin{equation}
    \pi(Y_i=1 \mid \mathbf{x}) = \psi_i = \frac{e^{\eta(\mathbf{x}_i)}}{1+e^{\eta(\mathbf{x}_i)}} = \frac{1}{1+e^{-\eta(\mathbf{x}_i)}}
    \label{eq:eq2.2}
\end{equation}
and its complement
\begin{equation}
    1 - \psi_i = \frac{e^{-\eta(\mathbf{x}_i)}}{1+e^{-\eta(\mathbf{x}_i)}} = \frac{1}{1+e^{\eta(\mathbf{x}_i)}}
    \nonumber
\end{equation}
where $\beta_k$ for $k = 1, \dotsc, p$ coefficients for covariates ($\mathbf{x}$). Hence, the logistic model is a binary classification or supervised learning algorithm. Equivalently, the model can be expressed using the logit function as follows
\begin{equation}
    \sigma(\psi_i) = \text{logit}(\psi_i) = \log \left(\frac{\psi_i}{1-\psi_i}\right) = \eta(\mathbf{x}_i)
    \label{eq:eq2.3}
\end{equation}
The sigmoid function $\sigma(\cdot)$ ensures predicted probabilities fall in the $\left[0, 1 \right]$ interval, compared to the linear probability model (LPM), which result in predicted probabilities out the $\left[0, 1 \right]$ interval. Alternatively, the probit link function, that is, the cumulative distribution function ($\Phi(\cdot)$) also yields the $\left[0, 1 \right]$ interval. However, the $\sigma(\cdot)$ is often preferred for its flexibility and in terms of interpretation of coefficient estimates. The likelihood of the data $Y_1, \dotsc, Y_n$ with respect to the intercept $\beta_0$ and coefficients $\beta_k$ is expressed as follows
\begin{equation}
    L(\beta_0, \beta_1 \dotsc, \beta_k \mid Y_1, \dotsc, Y_n) = \prod^n_{i=1}\psi_i^{Y_i}(1-\psi_i)^{1-Y_i}
    \label{eq:eq2.4}
\end{equation}
with corresponding log-likelihood
\begin{equation}
\begin{aligned}
    \ell(\boldsymbol{\beta}) &= \sum^{n}_{i = 1}Y_i\log(\psi_i) + (1 - Y_i)\log(1 - \psi_i) \\
    &= \sum^n_{i=1} Y_i\eta(\mathbf{x}_i) + \log(1-\psi_i)
    \end{aligned}
    \label{eq:eq2.5}
\end{equation}
From classical statistics, the Maximum Likelihood Estimates (MLEs) $\hat{\boldsymbol{\beta}}$ of $\boldsymbol{\beta}$ are obtained by taking the first derivative of the log-likelihood ($ \ell(\boldsymbol{\beta})$) with respect to $\boldsymbol{\beta}$ and setting it to zero, i.e.,  ($\partial\ell(\boldsymbol{\beta})/\partial \boldsymbol{\beta} = 0$). Then, iterative optimisation algorithm such as the Newton-Raphson algorithm, is used to find the MLEs, and confidence intervals are constructed following the notion of repeated sampling. Alternatively, Bayesian statistics can also be used, which allows for uncertainty quantification about the parameters $\boldsymbol{\beta} = \left(\beta_0, \dotsc, \beta_k \right)$. Prior beliefs about $\boldsymbol{\beta}$ are combined with information from data $\left[Y_i, \mathbf{x}_i\right]$ expressed by the likelihood in Equation~\ref{eq:eq2.4} to obtain the Bayesian posterior distribution $\pi(\boldsymbol{\beta} \mid \mathbf{Y}, \mathbf{x}) =  L(\boldsymbol{\beta} \mid Y_1, \dotsc, Y_n)\pi(\boldsymbol{\beta})$. The log of the posterior distribution is often used to avoid numerical problems (for example, overflow,  underflow, etc.). Thus, the log of the posterior distribution is as follows
\begin{equation}
    \log\pi(\boldsymbol{\beta} \mid \mathbf{Y}, \mathbf{x}) =  \ell(\boldsymbol{\beta})
    +  \sum_{k=1}^{p} \log \pi(\beta_k) 
    \label{eq:eq2.6}
\end{equation}
A Markov Chain Monte Carlo (MCMC) algorithm is designed and implemented to effectively explore the space of the posterior distribution to obtain samples for the parameters $\beta_0, \dotsc, \beta_k$ in the region of high posterior density. And finally, posterior summaries and 95\% posterior credible intervals can be obtained from the samples. There are already established software that allows for such posterior sampling such as the Windows version of Bayesian inference Using Gibbs Sampling (winBUS) \citep{winbugs2000}, Just Another Gibbs Sampler (JAGS) \citep{jags2004}, the Hamiltonian Monte Carlo algorithm implemented in the Stan \citep{neal2011}, with \textbf{R} package \textbf{cmdstan} \citep{cmdstan}, Numerical Inference for statistical Models using Bayesian and Likelihood Estimation (NIMBLE) \citep{deValpine2017}.



\subsection{Prevalence level}

Prevalence is the proportion of the population or the total number of observations $n$ with positive outcomes. The sum of all observations where $Y_{i} = 1$ in Equation~\ref{eq:eq2.1} is divided by the total number of observations to obtain the prevalence level. Since $Y_{i}$ is a random variable taking 1 or 0, the prevalence is the mean of $Y_{i}$, that is, $\bar{Y}_{i} = \frac{1}{n}\sum^n_{i=1}(Y_{i})$. The prevalence level ($\bar{Y}_{i}$) has enormous scientific importance in several fields; for example, in epidemiology it is often interpreted as how common a disease is in a population; also in ecology it is often referred to as the proportion of occupied area (POA), that is, the proportion of the study area that is occupied by a certain species of interest. The choice of the values of $\boldsymbol{\beta}$ has implications for prevalence; \cite{austin2023iterative} assumed the intercept determines the prevalence, which is true when $\boldsymbol{\beta} = \boldsymbol{0}$. However, in many real-world applications, this assumption is untenable. In a later section, the implications of this assumption will be demonstrated; thus, the need to relax it when generating data for the logistic regression model.
\section{Receiver operating characteristic curve (ROC)}\label{sec:roc}
It is important to assess the model classification accuracy or other classification algorithms. Classification accuracy refers to the ability of a model to discriminate between the positives and negatives \citep{janes2008}.
The ROC curve summarises the ability of a binary classifier to discriminate between two classes, for example, $Y_{i} = 1$ and $Y_{i}=0$. It is obtained by plotting the true positive rate (TPR), that is, the proportion of positives that are classified correctly, versus the false positive rate (FPR), that is, the proportion of negatives that are classified incorrectly, by varying the cutoff or threshold values. For instance, if the probability of the outcome variable, say $\mathbb{P}(Y \mid \mathbf{x}) > c$, is positive, it is classified as positive. On the other hand, if $\mathbb{P}(Y \mid \mathbf{x}) < c$ is classified as negative. This is done for different values of $c$ where $c$ is the cutoff value \citep{janes2008}. Therefore, changing the threshold value $c$ results in trade-offs between the TPR and FPR, and the ROC curve is used to provide a graphical view of these trade-offs \citep{lieli2019using}. The ROC curve lies in a unit square where the diagonal line passing through the coordinates \((0,0)\) and \((1,1)\) characterizes a poor classifier, and a curve that closely approaches the coordinate \((0, 1)\) represents a high-performance classifier \citep{dodd2003semiparametric}. 

The area under the receiver operating characteristic curve is analogous to the concordance statistic (C-statistic). The AUC value measures the probability that the classification algorithm assigns a higher probability to positive outcomes when they are positive \citep{pepe2004limitations,Danping2013}. It is the probability that, given a binary outcome with two classes (that is, positive and negative), the model assigns a higher probability to the positive class and a lower probability to the negative class \citep{pencina2008}. Take \(\hat{Y_{i}}\) as the predicted probabilities of the outcome variable $Y_i$, the AUC is interpreted as  \(\mathbb{P}(\hat{Y}_i > \hat{Y}_j)\) where the subscripts denote positive and negative outcomes. An AUC value closer to 0.5 implies that the classifier does not perform better than random chance, like flipping a fair coin. Whereas a value closer to 1 implies that the classifier has high discriminatory ability \citep{dodd2003semiparametric}. In practice, a good classifier will yield an AUC value closer to 1, the ideal value for a perfect model. 

\subsection{Relationship between AUC and covariates}

Covariate-specific AUC is the statistic used to measure the predictive accuracy of a classification algorithm or model \citep{Danping2013}. This implies that covariates have a bearing on the predictive accuracy of a model. Following \cite{janes2008} and \cite{Danping2013}, the relationship between the AUC and covariates can generally be expressed as follows;
\begin{equation}
     AUC = \text{logit}^{-1}(\mathbf{x}^\top \boldsymbol{\beta})
     \label{eq:eqn20}
\end{equation}
where $\text{logit}^{-1}$ is the logistic link function, $\boldsymbol{\beta}$ is a vector of parameters of interest \citep{Danping2013,dodd2003semiparametric}. The Equation~\ref{eq:eqn20} can be expanded further. Suppose we separate the covariates into $\mathbf{x}_i$ and $\mathbf{x}_j$, that is, the covariates for both the positive and negative observations. The predicted probabilities and the covariates for positive and negative observations can be denoted as $(\hat{Y}_i, \mathbf{x}_i)$ and $(\hat{Y}_j, \mathbf{x}_j)$, thus the AUC in this case can be written as $\mathbb{P}(\hat{Y}_i > \hat{Y}_j \mid \mathbf{x}_i, \mathbf{x}_j)$. Hence, the AUC compares the predicted probabilities $\hat{Y_i}$ for positive observations with its corresponding covariates $\mathbf{x}_i$ to the predicted probabilities $\hat{Y_i}$ for the negative observations with corresponding covariate $\mathbf{x}_j$ \citep{dodd2003semiparametric}. Hence, \cite{dodd2003semiparametric} proposed the following AUC regression model;
\begin{equation}
    \text{logit}(AUC) = \mathbf{x}^\top\boldsymbol{\beta}
    \label{eq:eqn21}
\end{equation}
This is analogous to Equation~\ref{eq:eqn20}, therefore when the logit link function is used, the AUC values can be interpreted in terms of odds. For regression models that utilize the logit link function (that is, $\text{logit(AUC)}=\mathbf{x}^\top\boldsymbol{\beta}$), the effect of the covariate is interpreted in terms of the odds. For example, an effect of a unit change of the covariate $\text{x}_1$ on the odds is $\exp(\beta_1)$ holding other covariates constant. Hence, the size of the $\boldsymbol{\beta}$ affects the model's accuracy. \cite{Danping2013} have derived a special case of Equation~\ref{eq:eqn20} and Equation~\ref{eq:eqn21} where the covariates $\mathbf{x}$ follow the standard normal distribution for both the positive and negative observations for the outcome. The AUC for this specific case is written as;
\begin{equation}
    AUC = \Phi \left(\frac{a}{\sqrt{b}} \right)
    \label{eq:eqn22}
\end{equation}
\begin{equation}
    AUC = \Phi \left(\frac{\mu_1-\mu_0}{\sqrt{\sigma^2_1+\sigma^2_0}} \right) 
     \label{eq:eqn23}
\end{equation}
where $\Phi$ is the standard normal cumulative distribution function, $a=(\mu_1 - \mu_0)$ and $b = \sigma^2_0 + \sigma^2_1$, $\mu_1$, $\mu_0$, $\sigma^2_1$ and $\sigma^2_0$ are the means and variances of $\mathbf{x}_i$ and $\mathbf{x}_j$ for $Y_i=1$ and $Y_i=0$, respectively. \cite{austin2012interpreting} derived another special case where the variances of $\mathbf{x}_i$ and $\mathbf{x}_j$ are equal ($\sigma^2_1 = \sigma^2_0$) Equation~\ref{eq:eqn23} is modified as follows;
\begin{equation}
    AUC = \Phi \left(\frac{\mu_1-\mu_0}{\sqrt{2}\sigma} \right) = \Phi \left(\frac{\frac{\sigma}{\sqrt{2}} (\mu_1-\mu_0)}{\frac{\sigma}{\sqrt{2}} \cdot \sqrt{2}\sigma} \right) = \Phi \left(\frac{\frac{\sigma}{\sqrt{2}} (\mu_1-\mu_0)}{\sigma^2} \right)  \nonumber
\end{equation}
\begin{equation}
    AUC = \Phi \left(\frac{\sigma}{\sqrt{2}} \cdot \frac{(\mu_1-\mu_0)}{\sigma^2} \right) = \Phi \left(\frac{\sigma \beta}{\sqrt{2}}\right)
    \label{eq:eqn24}
\end{equation}
where $\beta = (\mu_1-\mu_0)/\sigma^2$, hence, when the $\mathbf{x}$ is normally distributed with equal variance for $\mathbf{x}_i$ and $\mathbf{x}_j$ for both positive and negative observations, the AUC is a function of only the coefficient of $\mathbf{x}$ and the variance of $\mathbf{x}$ for both $Y_i = 1$ and $Y_i = 0$. We have seen from Equation~\ref{eq:eqn21} that the \text{logit}(AUC) is a function of $\boldsymbol{\beta}$ \citep{dodd2003semiparametric}. Equation~\ref{eq:eqn22} presents a special case where the AUC is a function of the difference in means of $\mathbf{x}_i$ and $\mathbf{x}_j$ denoted as $a$, and the sum of the variances of $\mathbf{x}_i$ and $\mathbf{x}_j$, denoted as $b$ \citep{Danping2013}. When we assume that the variance of $\mathbf{x}_i$ is equal to the variance of $\mathbf{x}_j$ \cite{austin2012interpreting}, the AUC is derived as a function of $\boldsymbol{\beta}$ and $\sigma$ using the standard cumulative normal distribution in Equation~\ref{eq:eqn24}. 
\section{Algorithms based on bisection method}\label{sec:algorithm}

The bisection method is a root-finding numerical algorithm based on the Intermediate Value Theorem \cite{Richard2011}. It is used to solve for the zeros of equations of the form $g(x) = 0$, for a given function $g$. If $f$ is a continuous function defined on an interval $(\theta_1, \theta_2)$, and $g(\theta_1)$ and $g(\theta_2)$ have opposite signs, the Intermediate Value Theorem says that there is some value $\theta$ within the interval $(\theta_1, \theta_2)$ for which $g(\theta) = 0$. \cite{Richard2011} provides a concise introduction to the bisection technique. The aim is to find the intercept and coefficient values that achieve a target AUC value and prevalence level. \cite{austin2023iterative} provides an iterative bisection method to explore this parameter space with two bisection algorithms; the first algorithm (Algorithm~\ref{alg:algorithm1}) finds the intercept value that yields some target prevalence level. The second algorithm (Algorithm~\ref{alg:algorithm2}) finds the target AUC value using the intercept value obtained from Algorithm~\ref{alg:algorithm1}. Algorithm~\ref{alg:algorithm1} and Algorithm~\ref{alg:algorithm2} do not depend on each other sequentially to find the target prevalence level. The details of the implementation of these algorithms are presented in \citep{austin2023iterative}.

Algorithm~\ref{alg:algorithm1} and Algorithm~\ref{alg:algorithm2}, as proposed by \cite{austin2023iterative} to find the target prevalence and predictive performance (AUC), respectively, are presented. Throughout, \(\textbf{x}\) represent covariates, $\boldsymbol{\beta}$ represent coefficients of the covariates, \(\beta_\textit{init}\) is initial coefficients values, \(n\) is sample size, target prevalence ($\tau_{\text{prev}}$), observed prevalence ($\hat{p}$), target AUC ($\tau_{\text{auc}}$), empirical AUC ($\hat{C}$), tolerance level of prevalence ($\epsilon_{\text{prev}}$) and tolerance level of AUC, ($\epsilon_{\text{auc}})$.
\LinesNotNumbered
\begin{algorithm}[!htbp]
\small
\caption{Determine intercept $\beta_0$ for target prevalence}
\KwIn{$\mathbf{x}$, $\boldsymbol{\beta}$, $\tau_{\text{prev}}$,
      $[\beta_0^{\text{low}},\, \beta_0^{\text{high}}]$,
      $\epsilon_{\text{prev}}$}
\KwOut{Intercept $\beta_0$}
\While{$|\hat{\psi} - \tau_{\text{prev}}| > \epsilon_{\text{prev}}$}{
    $\beta_0 \leftarrow (\beta_0^{\text{low}} + \beta_0^{\text{high}}) / 2$\\ 
    \texttt{Compute predicted probabilities:} $\psi_i \leftarrow \left[1 + \exp\!\left(-(\beta_0 + \mathbf{x}_i^{\top}\boldsymbol{\beta})\right)\right]^{-1}$\\
     \texttt{Estimated prevalence:} $\hat{\psi} \leftarrow n^{-1}\sum_{i=1}^{n} \psi_i$\\
    \eIf{$\hat{\psi} < \tau_{\text{prev}}$}{
        $\beta_0^{\text{low}} \leftarrow \beta_0$\\
    }{
        $\beta_0^{\text{high}} \leftarrow \beta_0$
    }
}
\label{alg:algorithm1}
\end{algorithm}
\begin{algorithm}[!htbp]
\small
\caption{Determine scalar $\sigma$ for target AUC value}
\KwIn{$\mathbf{x}$, $\boldsymbol{\beta}_{\text{init}}$,
       $\beta_0$, $\tau_{\text{auc}}$, $[\sigma^{\text{low}},\, \sigma^{\text{high}}]$, $\epsilon_{\text{auc}}$}
\KwOut{Scalar $\sigma$ and scaled coefficients $\boldsymbol{\beta}_{\sigma}$}
\texttt{Initial AUC value:} $\hat{C} \leftarrow 0$\;
\While{$|\hat{C} - \tau_{\text{auc}}| > \epsilon_{\text{auc}}$}{
    $\sigma \leftarrow (\sigma^{\text{low}} + \sigma^{\text{high}}) / 2$\\
    \texttt{Scaled coefficients:} $\boldsymbol{\beta}_{\sigma} \leftarrow \sigma \cdot \boldsymbol{\beta}_{\text{init}}$\\ 
     \texttt{Compute predicted probabilities:} $\psi_i \leftarrow \left[1 + \exp\!\left(-(\beta_0 + \mathbf{x}_i^{\top}\boldsymbol{\beta}_{\sigma})\right)\right]^{-1}$\\
    \texttt{Sample outcomes:} $Y_i \sim \text{Bernoulli}(p_i)$\\
    \texttt{Extract the empirical AUC value:} $\hat{C} \leftarrow \text{AUC}(Y,\,p)$\\
    \eIf{$\hat{C} < \tau_{\text{auc}}$}{
        $\sigma^{\text{low}} \leftarrow \sigma$\\
    }{
        $\sigma^{\text{high}} \leftarrow \sigma$
    }
   }
   
\label{alg:algorithm2}
\end{algorithm}
\DontPrintSemicolon
\begin{algorithm}[!htbp]
\small
\caption{Integrated bisection method}
\label{alg:bisection}
\KwIn{$\mathbf{x}$, $\boldsymbol{\beta}_{\text{init}}$,
      $\tau_{\text{prev}}$, $\tau_{\text{auc}}$, $\epsilon_{\text{prev}}$, $\epsilon_{\text{auc}}$}
\KwOut{Intercept $\beta_0$, scalar $\sigma$, scaled coefficients $\boldsymbol{\beta}_{\sigma}$}
\Repeat{$|\hat{\psi} - \tau_{\text{prev}}| \leq \epsilon_{\text{prev}}$
        \normalfont\textbf{and}
        $|\hat{C} - \tau_{\text{auc}}| \leq \epsilon_{\text{auc}}$}{
    \texttt{Find $\beta_0$ via Algorithm~1} \\
    $\beta_0 \leftarrow \textsc{FindIntercept}\!\left(\mathbf{x},\,
                         \boldsymbol{\beta}_{\sigma},\, \tau_{\text{prev}},\,
                         \epsilon_{\text{prev}}\right)$\\
                         
    \texttt{Find $\sigma$ via Algorithm~2} \\
    $(\sigma,\, \boldsymbol{\beta}_{\sigma}) \leftarrow
        \textsc{FindScalar}\!\left(\beta_0,\, \mathbf{x},\,
                            \boldsymbol{\beta}_{\text{init}},\, \tau_{\text{auc}},\,
                            \epsilon_{\text{auc}}\right)$}
\label{alg:algorithm3}
\end{algorithm}

\section{Simulation}\label{sec:sim}

Using a sample size of $n = 5000$, we consider two simulation scenarios. In both scenarios, the target AUC is fixed at $\tau_{\text{auc}} = 0.85$, while the target prevalence is set to $\tau_{\text{prev}} = 0.10$ and $\tau_{\text{prev}} = 0.80$, respectively. The tolerance levels are $\epsilon_{\text{prev}} = 0.0001$ for the prevalence and $\epsilon_{\text{auc}} = 0.0001$ for the AUC. Algorithms~\ref{alg:algorithm1}, \ref{alg:algorithm2}, and \ref{alg:algorithm3} are each run $m=100$ times. To assess how well an algorithm simultaneously achieves the target prevalence and target AUC, we compute the mean joint squared distance from the target values for simulation run $i$, \[ \bar d = \frac{1}{m}\sum_{i=1}^{m} [(\hat{p}_i-\tau_{\text{prev}})^2 + (\hat{C}_i-\tau_{\text{auc}})^2] \] where $\hat{p}_i$ and $\hat{C}_i$ denote the estimated prevalence and AUC, respectively, from the $i$th simulation run. Smaller values of $\bar d$ indicate better agreement with the target prevalence and target AUC, and therefore better overall performance of the algorithm. Alternatively, the mean of the square root of the joint distance, $\sqrt{d}$, can also be used. In particular, when the mean joint distance is close to zero, the mean of $\sqrt{d}$ provides a more informative measure for comparing the three algorithms.
\begin{figure}[htbp]
    \centering
    \begin{subfigure}[t]{0.7\textwidth}
        \centering
        \includegraphics[width=\textwidth]{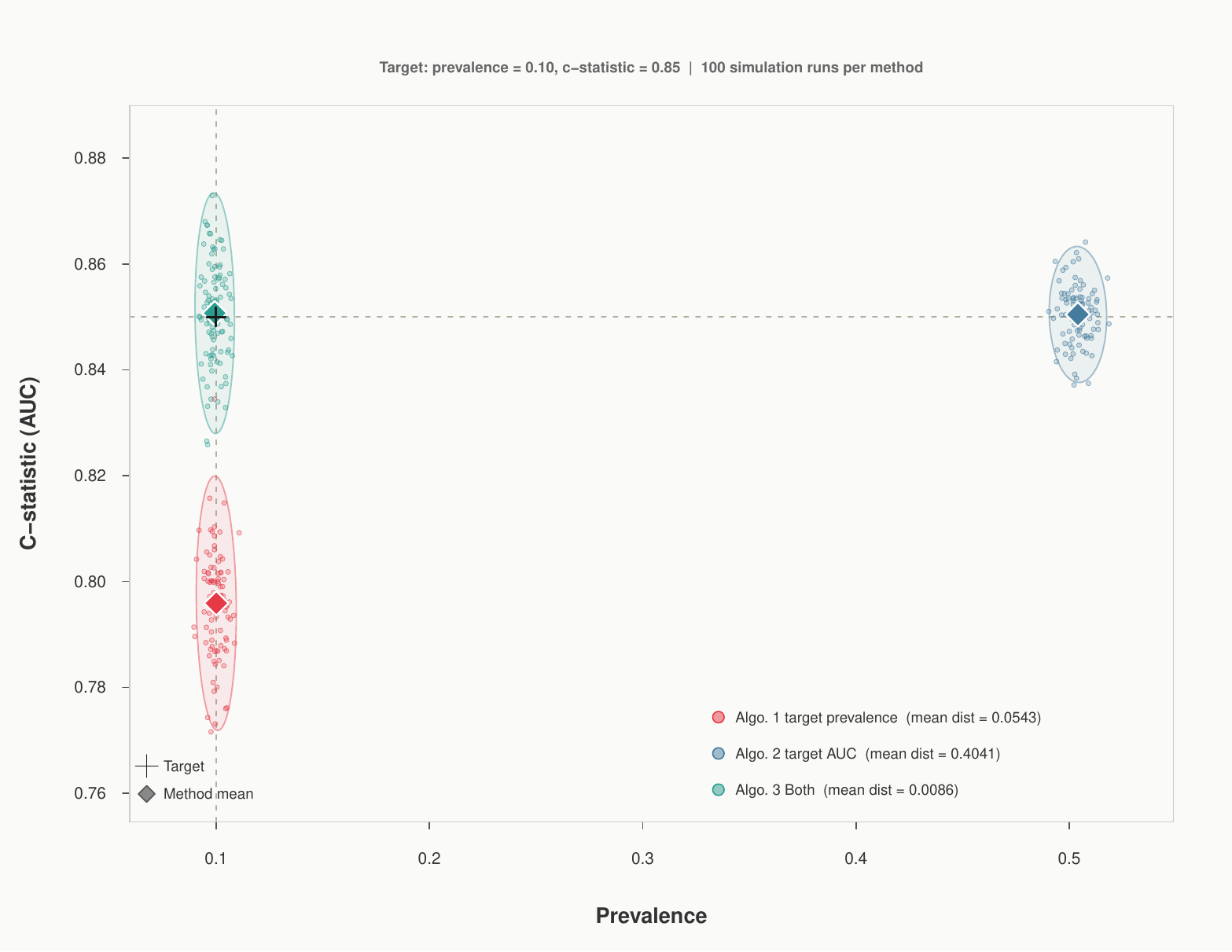}
        \caption{Target prevalence of 0.10 and target C-statistic (AUC) of O.85.}
        \label{fig:algo1}
    \end{subfigure}
    \hfill
    \begin{subfigure}[t]{0.7\textwidth}
        \centering
        \includegraphics[width=\textwidth]{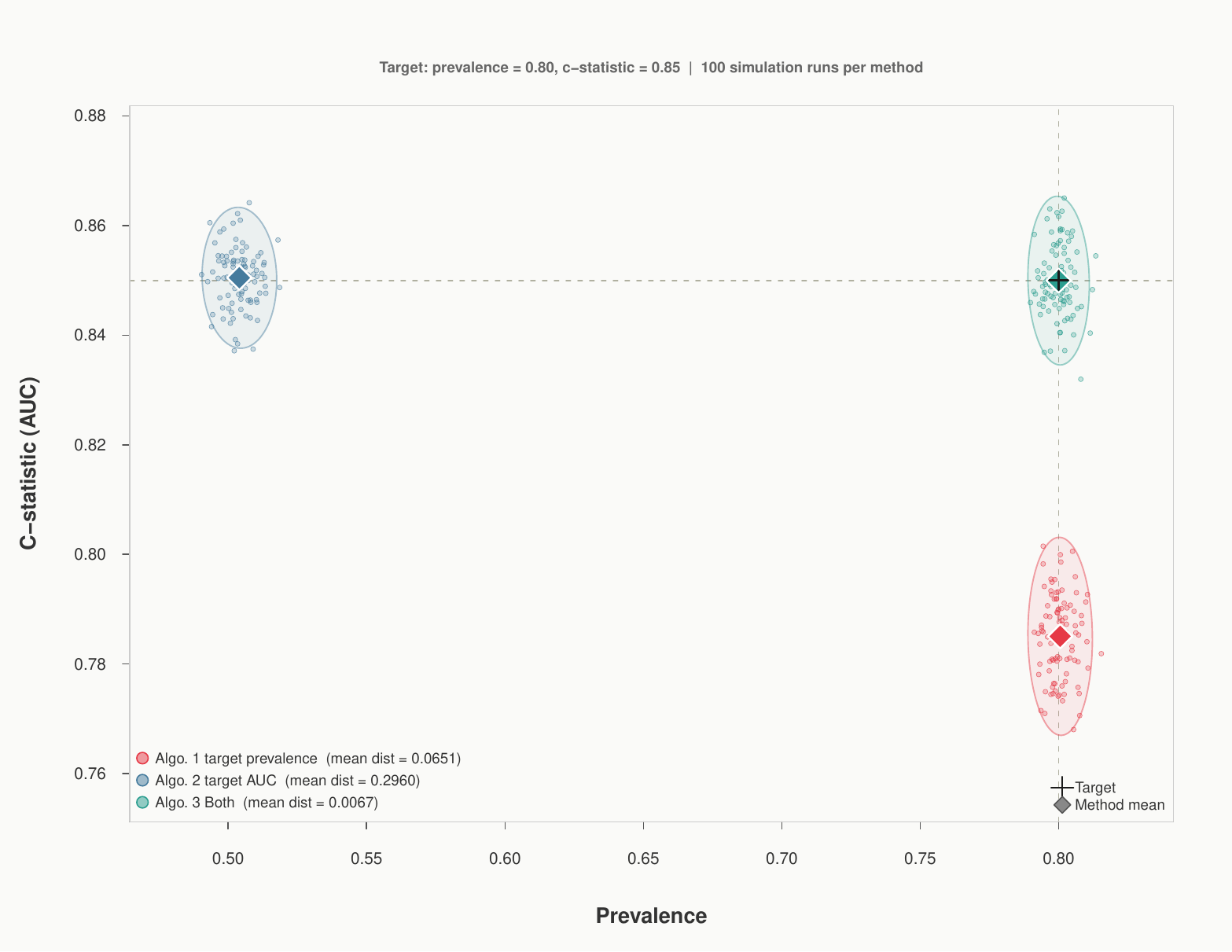}
        \caption{Target prevalence of 0.80 and target C-statistic (AUC) of 0.85.}
        \label{fig:algo2}
    \end{subfigure}
    \caption{Results from 100 simulation runs for the three algorithms. Each point represents the prevalence and AUC estimated from a simulated dataset, while the diamonds indicate the mean values across runs. The dashed vertical and horizontal lines denote the target prevalence and target AUC, respectively. Algorithm~1 targets the prevalence, Algorithm~2 targets the AUC, and Algorithm~3 targets both quantities simultaneously. Panel~(a) corresponds to a target prevalence of 0.10 and target AUC of 0.85, whereas panel~(b) corresponds to a target prevalence of 0.80 and target AUC of 0.85.}
    \label{fig:compare-algo}
\end{figure}

In Figure~\ref{fig:algo1}, the target values are $\tau_{\text{prev}}=0.10$ and $\tau_{\text{auc}}=0.85$. Algorithm~\ref{alg:algorithm1} (Algo.~1) identifies the intercept ($\beta_0$) that achieves the target prevalence, but the resulting estimated AUC, $\hat{C}$, falls below the target value. Conversely, Algorithm~\ref{alg:algorithm2} (Algo.~2) attains the target AUC but fails to achieve the desired prevalence, producing a prevalence greater than 0.10. Thus, Algorithms~\ref{alg:algorithm1} and \ref{alg:algorithm2} can each achieve only one of the two targets. In contrast, Algorithm~\ref{alg:algorithm3} (Algo.~3) successfully achieves both the target prevalence and target AUC. 

A similar pattern is observed in Figure~\ref{fig:algo2}, where $\tau_{\text{prev}}=0.80$ and $\tau_{\text{auc}}=0.85$. Again, Algorithms~\ref{alg:algorithm1} and \ref{alg:algorithm2} are able to achieve only one of the target quantities, whereas Algorithm~\ref{alg:algorithm3} achieves both simultaneously. 

This superior performance is reflected in the mean joint distance, $\bar d$. In both scenarios, Algorithm~\ref{alg:algorithm3} yields the smallest mean joint distance, with $\bar d = 0.009$ and $\bar d = 0.007$ for Figures~\ref{fig:algo1} and \ref{fig:algo2}, respectively. These values are substantially lower than those obtained by Algorithm~\ref{alg:algorithm1} ($\bar d = 0.054$ and $\bar d = 0.065$) and Algorithm~\ref{alg:algorithm2} ($\bar d = 0.404$ and $\bar d = 0.296$).

\section{Distribution of predicted probabilities}\label{sec:dist_predicted_prob}

\begin{figure}[htbp]
    \centering
        \includegraphics[width=0.8\textwidth]{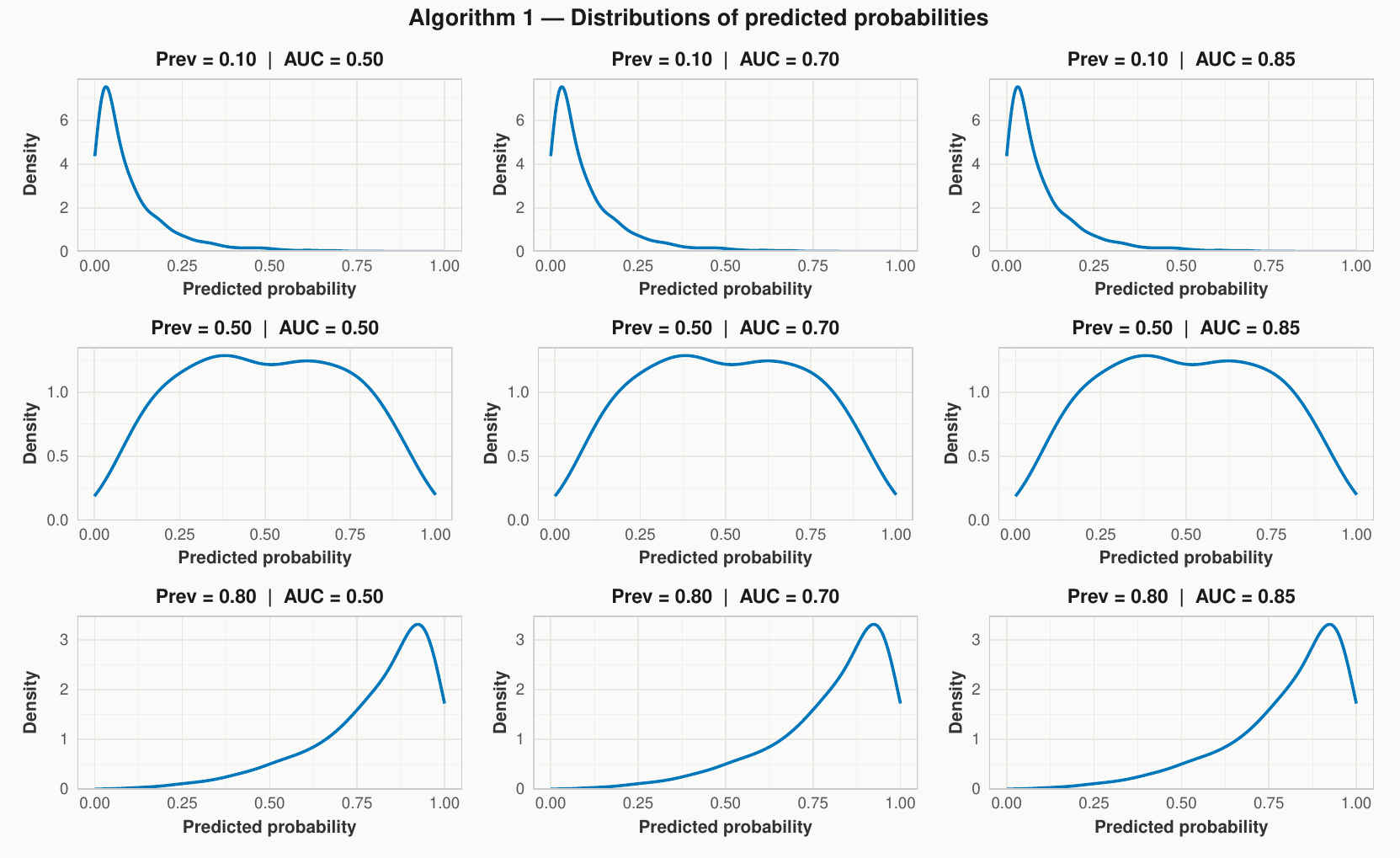}
\caption{Distributions of predicted probabilities obtained using Algorithm~\ref{alg:algorithm1}. Rows correspond to target prevalence values $\{0.10, 0.50, 0.80\}$ and columns correspond to target AUC values $\{0.50, 0.70, 0.85\}$. The distributions remain unchanged across columns for a fixed prevalence, demonstrating that Algorithm~\ref{alg:algorithm1} is determined solely by the target prevalence and is unaffected by the target AUC.}        \label{fig:algo1predictedprob}
\end{figure}

\begin{figure}[htbp]
    \centering
        \includegraphics[width=0.8\textwidth]{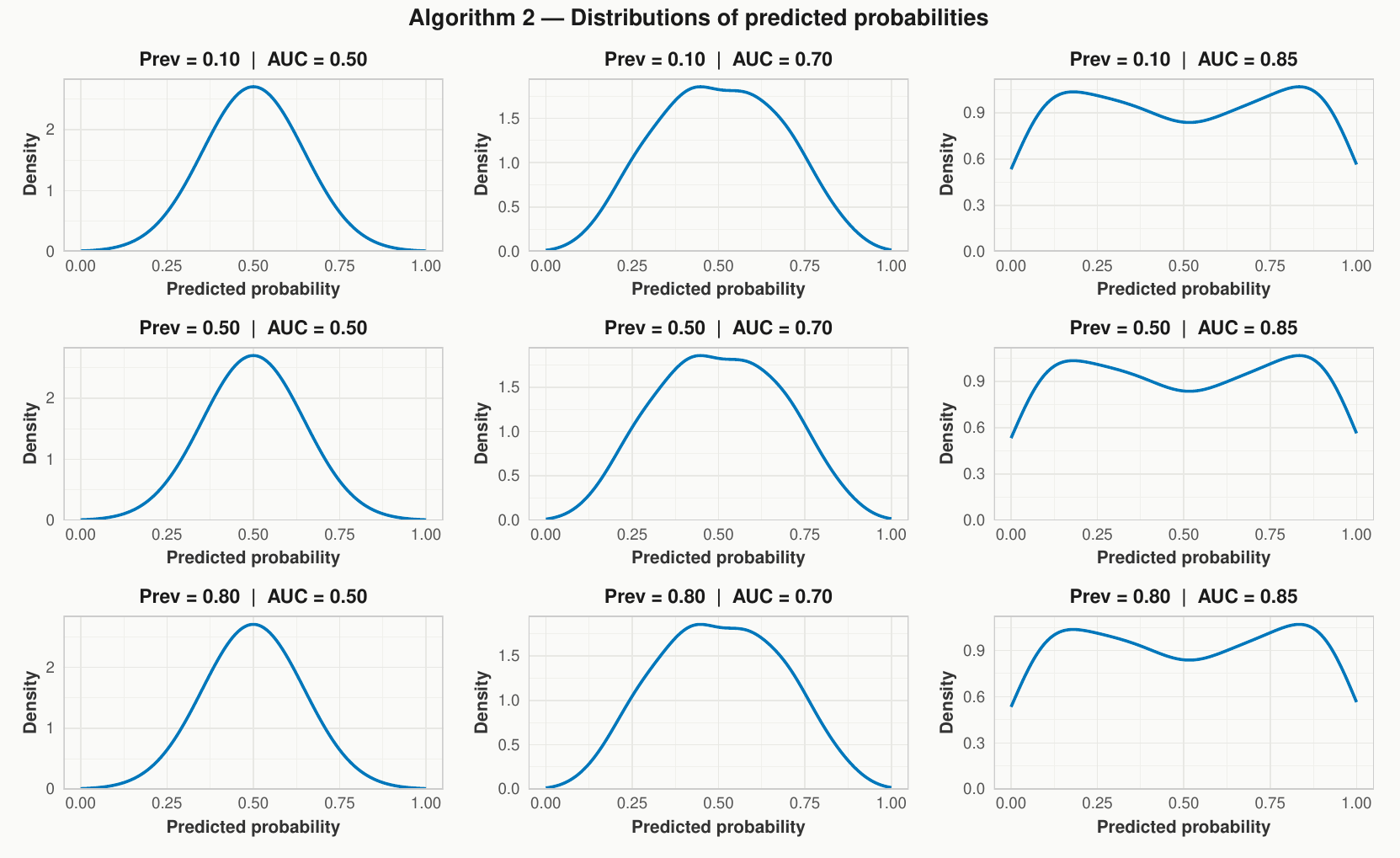}
        \caption{Distributions of predicted probabilities obtained using Algorithm~\ref{alg:algorithm2}. Rows correspond to target prevalence values $\{0.10, 0.50, 0.80\}$ and columns correspond to target AUC values $\{0.50, 0.70, 0.85\}$. The distributions remain unchanged across rows for a fixed AUC, demonstrating that Algorithm~\ref{alg:algorithm2} is determined solely by the target AUC and is unaffected by the target prevalence.}
        \label{fig:algo2predictedprob}
\end{figure}

\begin{figure}[htbp]
    \centering
        \includegraphics[width=0.8\textwidth]{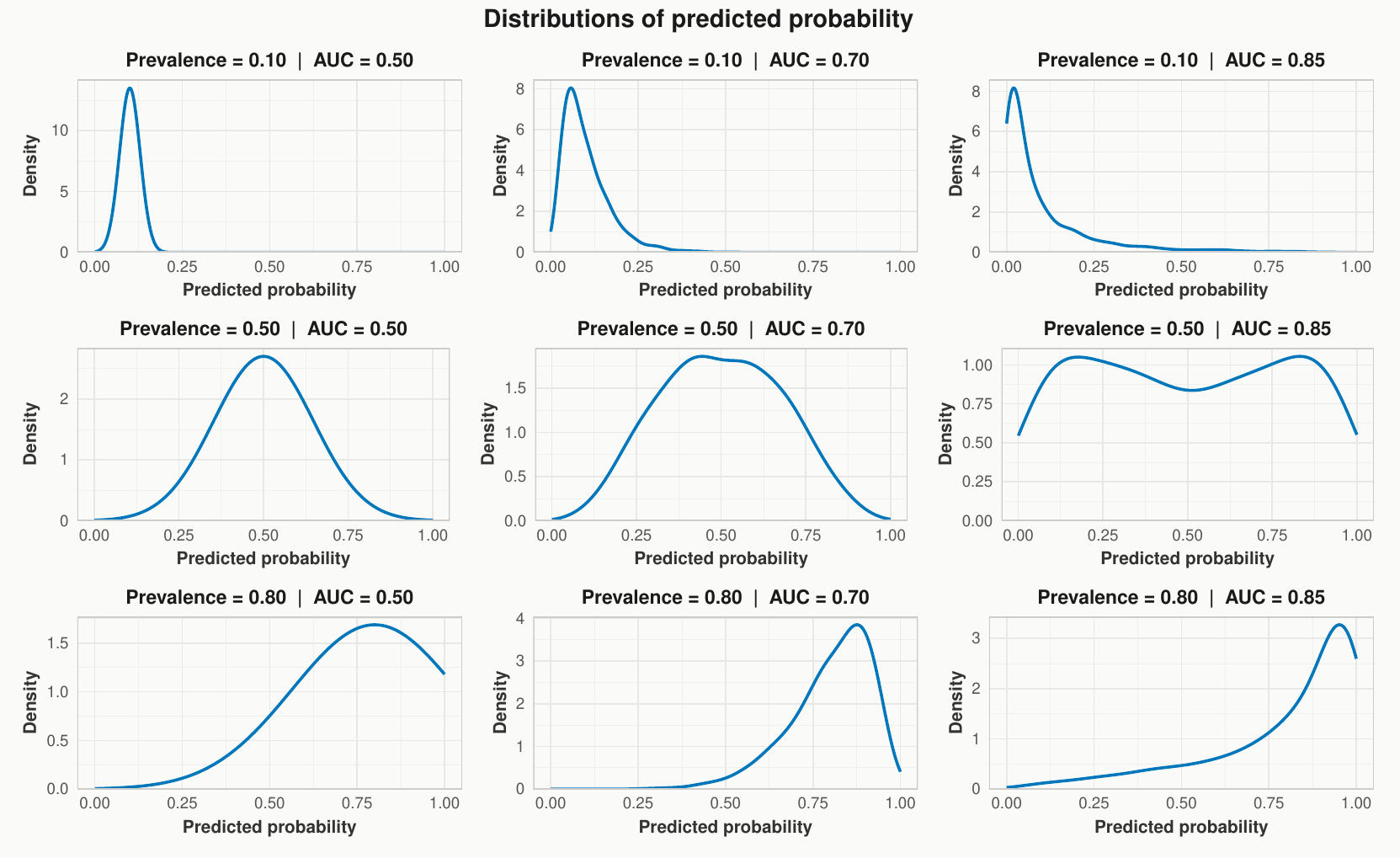}
\caption{Distributions of predicted probabilities obtained using Algorithm~\ref{alg:algorithm3}. Rows correspond to target prevalence values $\{0.10, 0.50, 0.80\}$ and columns correspond to target AUC values $\{0.50, 0.70, 0.85\}$. Changes in both prevalence and AUC alter the shape of the distributions, demonstrating that Algorithm~\ref{alg:algorithm3} simultaneously incorporates both target characteristics.}
        \label{fig:predictedprob}
\end{figure}

Figures~\ref{fig:algo1predictedprob}, \ref{fig:algo2predictedprob}, and \ref{fig:predictedprob} illustrate the distributions of the predicted probabilities obtained under different combinations of target prevalence and target AUC values. Because Algorithm~\ref{alg:algorithm1} is designed to achieve only a target prevalence, the resulting distributions are determined solely by the prevalence. Similarly, Algorithm~\ref{alg:algorithm2} is designed to achieve only a target AUC, and therefore the distributions depend only on the specified AUC value. In contrast, Algorithm~\ref{alg:algorithm3} simultaneously targets both prevalence and AUC, producing distributions whose shape reflects both characteristics of the data-generating process.

In Figure~\ref{fig:algo1predictedprob} and \ref{fig:algo2predictedprob}, the distributions of the predicted probabilities for the varying prevalence and AUC values are presented. Algorithm~\ref{alg:algorithm1} and Algorithm~\ref{alg:algorithm2}, which are designed only to find target prevalence and target AUC separately, yield predicted distributions that are influenced by only one of the desired target. However, Algorithm~\ref{alg:algorithm3} yields predicted distributions that are influenced by both desired properties of the data. 


\begin{figure}[htbp]
    \centering
    \includegraphics[width=0.8\textwidth]{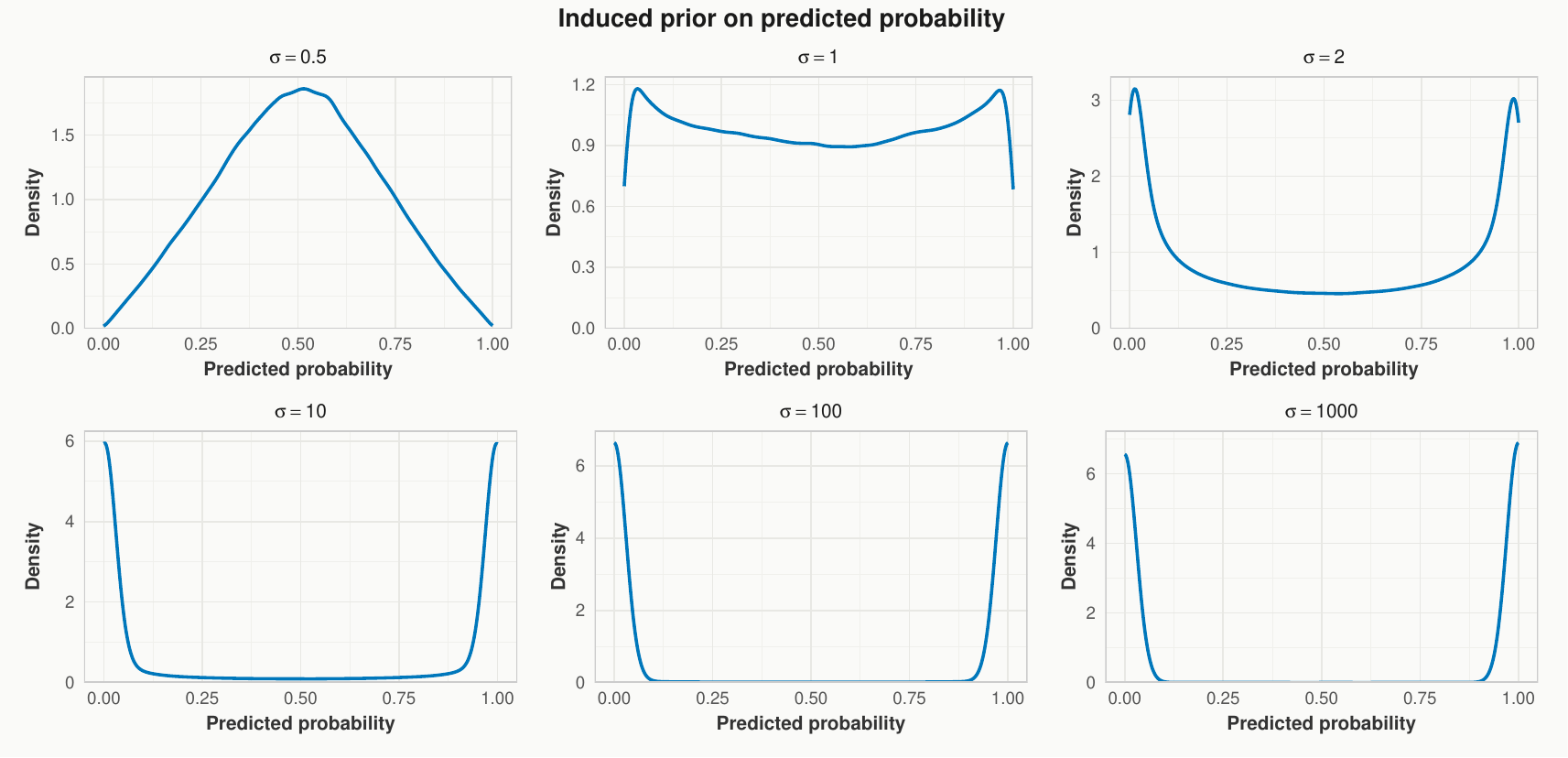}
        \caption{The induced priors at the different values of standard deviations for
        the intercept ($\beta_0$) and coefficients ($\boldsymbol{\beta}$). The choice 
        of vague prior distributions for the $\beta_0$ and $\boldsymbol{\beta}$ does 
        not induce a vague prior on the distribution of the predicted probability. A 
        smaller standard deviation ($\sigma$), which implies an informative prior on 
        $\beta_0$ and $\boldsymbol{\beta}$ induces meaningful predicted probabilities compared to larger standard deviations.}
        \label{fig:inducedpriors}
\end{figure}

Data simulation for Bayesian hierarchical models has been extensively investigated, for example, \cite{joseph2018comment},  \cite{banner2020use}, and \cite{newman2025logistic}. As indicated in Section~\ref{sec:logistic}, the posterior distribution in Equation~\ref{eq:eq2.6} has two components, that is, the likelihood and priors for $\beta_0$ and $\beta_k$ for $k = 1, \dotsi p$. To allow for the dominance of the data expressed by the likelihood in the posterior, vague prior distributions are often employed for $\beta_0$ and $\beta_k$. Figure~\ref{fig:inducedpriors} illustrates the induced prior distributions of the predicted probabilities arising from different prior specifications for the intercept and regression coefficients. Although the priors are specified on the log-odds scale through $\beta_0$ and $\beta_k$, they induce non-trivial prior distributions on the probability scale via the logistic transformation. Consequently, seemingly vague priors on the regression parameters do not necessarily correspond to vague prior beliefs about the predicted probabilities. Thus, whereas the priors for $\beta_0$ and $\beta_k$ are on the log-odds scale, the predicted values are on the probability scale ($\left[0, 1\right]$) obtained via the logistic transformation.

This phenomenon has important implications in practice. For example, in occupancy models and species distribution models (SDMs), $\psi_i$ is interpreted as the occupancy probability. The mean of $\psi_i$, referred to as prevalence in Section~\ref{sec:logistic}, is commonly interpreted as the proportion of area occupied (PAO) by a species. Similarly, in disease modelling and epidemiology, prevalence corresponds to the proportion of the population affected by a disease. Consequently, prior specification for the intercept and regression coefficients requires careful consideration, since apparently vague priors may induce strong or unrealistic prior beliefs on the probability scale.

There are also potential numerical consequences of highly diffuse priors. As shown in Figure~\ref{fig:inducedpriors}, when the prior standard deviation becomes very large (e.g., $\sigma=1000$), the linear predictor in Equation~\ref{eq:eq2.3} can take extremely large positive or negative values, leading to predicted probabilities that are numerically indistinguishable from 0 or 1. 

Figure~\ref{fig:inducedpriors} also highlights the relationship between the spread of the regression coefficients and the discriminatory ability of the model. As $\sigma$ increases, the predicted probabilities become increasingly concentrated near 0 and 1, indicating near-perfect separation between the positive and negative classes. A similar phenomenon is observed when Algorithms~\ref{alg:algorithm2} and \ref{alg:algorithm3} are used to generate data with target AUC values approaching 1. In both cases, the effective magnitude of the regression coefficients increases, producing increasingly extreme linear predictors and hence probabilities close to 0 and 1. Therefore, highly diffuse priors in Bayesian logistic regression can induce prior predictive behaviour analogous to that of a data-generating mechanism with near-perfect classification performance.

\begin{figure}[htbp]
    \centering
    \includegraphics[width=0.8\textwidth]{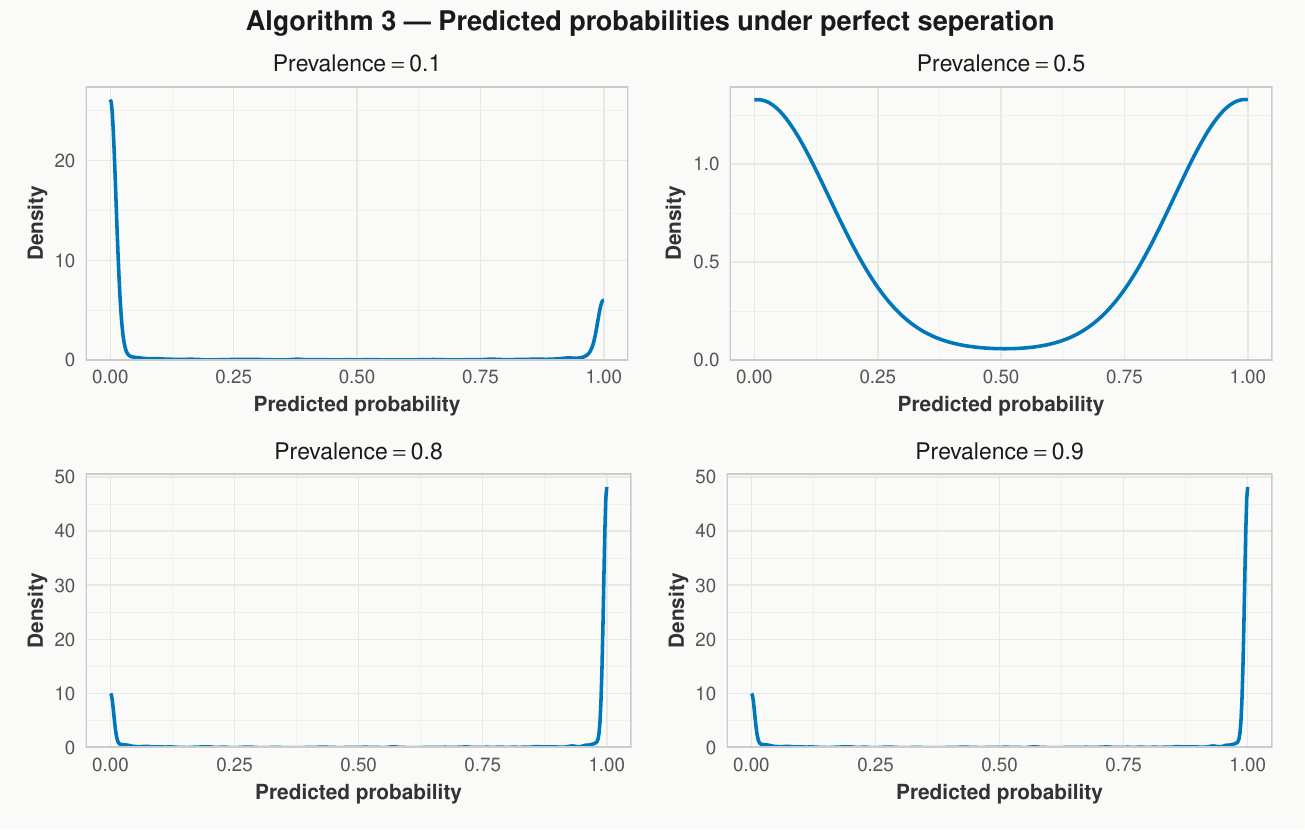}
        \caption{Induced prior distributions of the predicted probabilities under different prior standard deviations for the intercept ($\beta_0$) and coefficients ($\boldsymbol{\beta}$). Increasing the prior variance on the regression parameters does not result in increasingly diffuse distributions for the predicted probabilities. Instead, large standard deviations place substantial prior mass near 0 and 1, whereas smaller values of $\sigma$ yield more concentrated and interpretable prior predictive distributions.}
        \label{fig:perfectsep}
\end{figure}
\section{Discussion}\label{sec:disc} Logistic regression is widely used across disciplines such as economics, ecology, engineering, and epidemiology. In many applications, both prevalence and discriminatory ability, commonly measured by the AUC, are important characteristics of the data. However, when simulating data from a logistic regression model, controlling these quantities simultaneously is not straightforward. 

Section~\ref{sec:roc} established the relationship between the AUC and the regression coefficients under the assumption of normally distributed covariates. The proposed Algorithm~\ref{alg:algorithm3}, which combines Algorithms~\ref{alg:algorithm1} and \ref{alg:algorithm2}, does not rely on this assumption to achieve specified target prevalence and AUC values. Simulation studies show that Algorithm~\ref{alg:algorithm3} consistently produces smaller mean joint distances than Algorithms~\ref{alg:algorithm1} and \ref{alg:algorithm2}, including in settings with binary covariates (see Figure~\ref{fig:compare-algo-categorical} in the Supplementary Material). Although performance remains good in the presence of categorical covariates, the mean joint distance is larger than in the continuous-covariate setting. 

The use of vague priors is common in Bayesian applications because it avoids imposing strong subjective assumptions \citep{banner2020use}. However, for logistic regression, diffuse priors on the intercept and regression coefficients can induce highly informative prior beliefs on the probability scale. As demonstrated in Figure~\ref{fig:inducedpriors}, large prior standard deviations produce predicted probabilities concentrated near 0 and 1, corresponding to near-perfect separation and AUC values approaching one. Such behaviour may also lead to numerical difficulties when fitting logistic regression models.

To address this issue, \cite{newman2025logistic} recommended standard normal priors for regression coefficients when covariates are standardised, together with special treatment of the intercept. Their focus, however, was on the prior predictive distribution rather than on achieving target prevalence and predictive performance. Although Algorithm~\ref{alg:algorithm3} generates data for classical logistic regression and produces point estimates of $\beta_0$ and $\boldsymbol{\beta}$ rather than full posterior distributions, it can be used alongside such prior-specification guidelines to generate data with desired prevalence and discrimination properties for Bayesian applications.

\bibliographystyle{apalike}
\bibliography{references}

\renewcommand{\thefigure}{S\arabic{figure}}
\setcounter{figure}{0}
\appendix
\clearpage
\begin{center}
    {\LARGE \textbf{Supplementary Material}}
\end{center}
\vspace{1em}

\begin{figure}[H]
    \centering
    \begin{subfigure}[t]{0.7\textwidth}
        \centering
        \includegraphics[width=\textwidth]{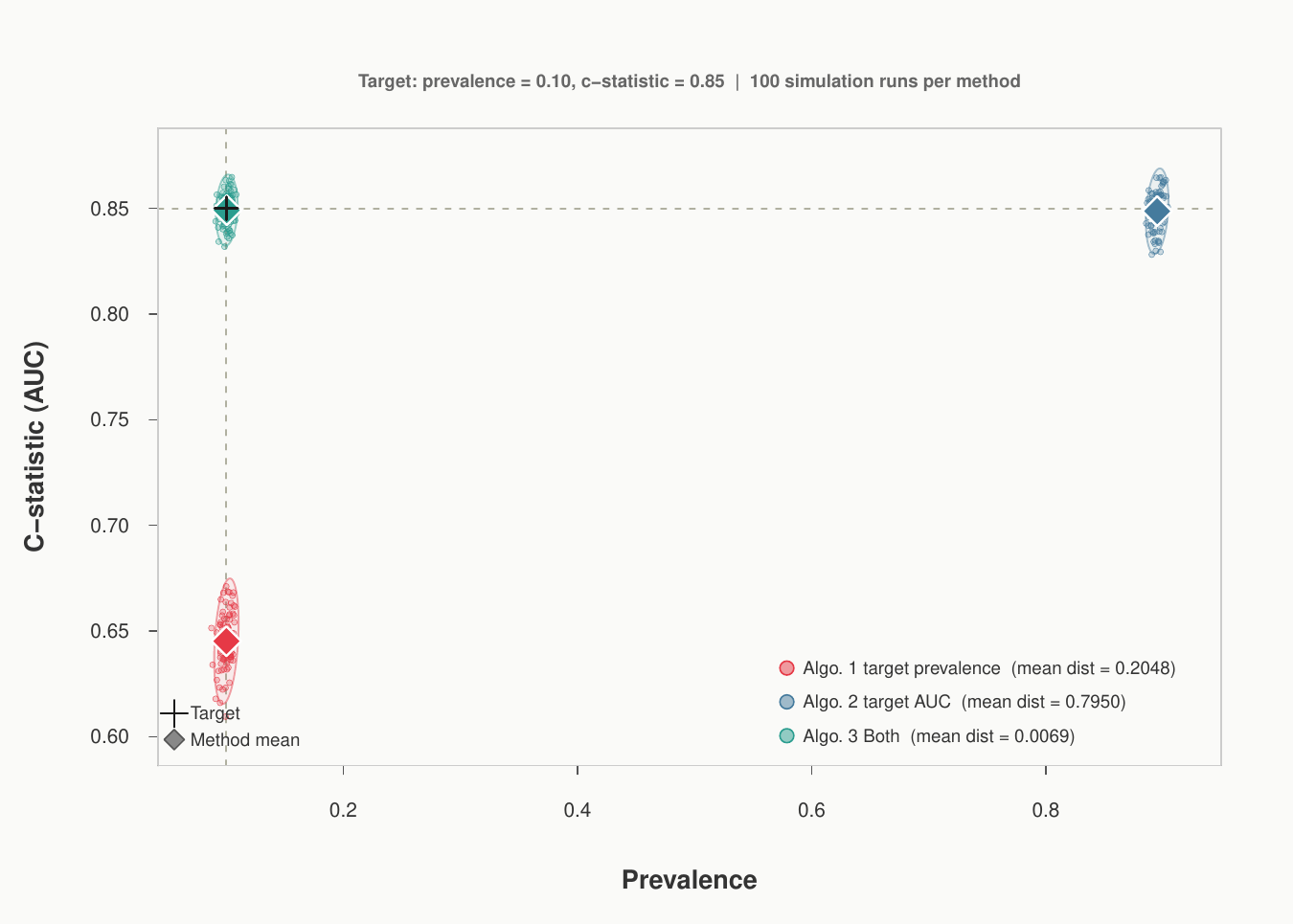}
        \caption{Target prevalence of 0.10 and target C-statistic (AUC) of O.85.}
        \label{fig:algocat1}
    \end{subfigure}
    \hfill
    \begin{subfigure}[t]{0.7\textwidth}
        \centering
        \includegraphics[width=\textwidth]{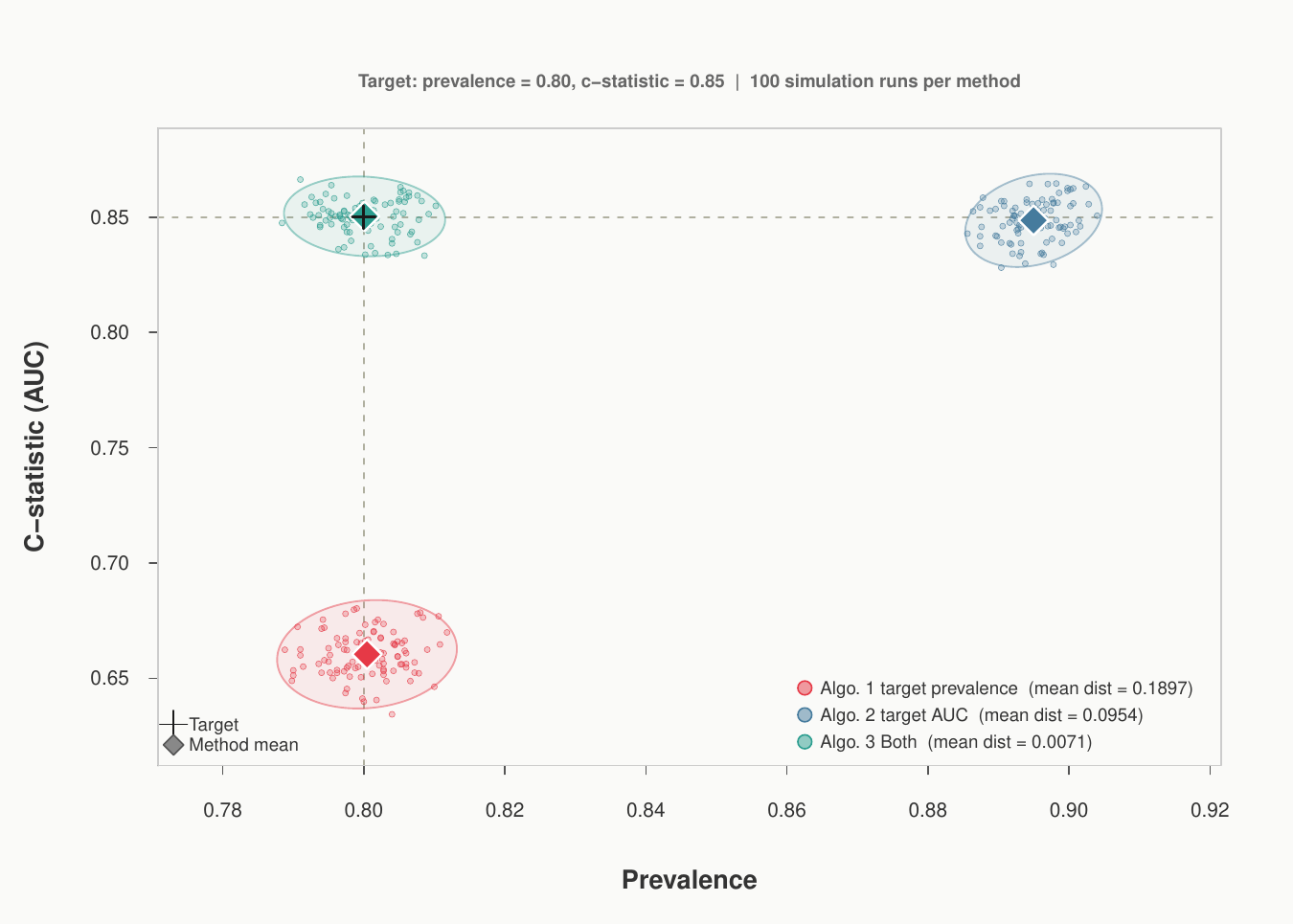}
        \caption{Target prevalence of 0.80 and target C-statistic (AUC) of 0.85.}
        \label{fig:algocat2}
    \end{subfigure}
    \caption{
    For 100 simulation runs, the results of the two algorithms, that is, Algorithm~1 to find the intercept that yields a target prevalence, Algorithm~2 that yields a target AUC value, and Algorithm~3 that achieves both target prevalence and target AUC values. Here we set the target prevalence to 0.10 for (a), and 0.80 for (b), and target C-statistic (AUC) to 0.85 for both (a) and (b).}
    \label{fig:compare-algo-categorical}
\end{figure}

\end{document}